\documentclass[final,3p,times,twocolumn]{elsarticle}

\usepackage{color, graphicx, dcolumn, bm, multirow}
\usepackage[mathlines]{lineno}

\usepackage{stfloats}
\usepackage{subfigure}
\usepackage{subcaption}
\usepackage{amsmath}
\usepackage{float}
\usepackage{indentfirst}
\usepackage[utf8]{inputenc}
\usepackage{listings}
\usepackage{algorithm}
\usepackage{xcolor}
\usepackage{appendix}
\usepackage{mathrsfs}
\usepackage{siunitx}
\usepackage{soul}
\usepackage[normalem]{ulem}
\usepackage{booktabs}
\usepackage[colorlinks=true,allcolors=blue]{hyperref}
\usepackage{enumitem}
\usepackage{pifont}
\usepackage{amssymb}
\usepackage{amsmath}
\usepackage{amsthm}

\begin{document}

\begin{frontmatter}

\title{Searching for Neutrinoless Double-Beta Decay of $^{136}$Xe with PandaX-4T}

\affiliation[TDLee]{organization={New Cornerstone Science Laboratory, Tsung-Dao Lee Institute, Shanghai Jiao Tong University},city={Shanghai},postcode={201210},country={China}}
\affiliation[SYU]{organization={School of Physics, Sun Yat-sen University},city={Guangzhou},postcode={510275},country={China}}
\affiliation[shKeyLab]{organization={School of Physics and Astronomy, Shanghai Jiao Tong University, Key Laboratory for Particle Astrophysics and Cosmology (MoE), Shanghai Key Laboratory for Particle Physics and Cosmology},city={Shanghai},postcode={200240},country={China}}
\affiliation[SJTUSC]{organization={Shanghai Jiao Tong University Sichuan Research Institute},city={Chengdu},postcode={610213},country={China}}
\affiliation[scKeyLab]{organization={Jinping Deep Underground Frontier Science and Dark Matter Key Laboratory of Sichuan Province},city={Xichang},postcode={615000},country={China}}

\affiliation[BUAA]{organization={School of Physics, Beihang University},city={Beijing},postcode={102206},country={China}}

\affiliation[BUAACenter]{
    organization={Peng Huanwu Collaborative Center for Research and Education, Beihang University},
    city={Beijing},
    postcode={100191},
    country={China}}

\affiliation[BUAALab]{
    organization={Beijing Key Laboratory of Advanced Nuclear Materials and Physics, Beihang University},
    city={Beijing},
    postcode={102206},
    country={China}}

\affiliation[SCNT]{
    organization={Southern Center for Nuclear-Science Theory (SCNT), Institute of Modern Physics, Chinese Academy of Sciences},
    city={Huizhou},
    postcode={516000},
    country={China}}

\affiliation[USTClab]{
    organization={State Key Laboratory of Particle Detection and Electronics, University of Science and Technology of China},
    city={Hefei},
    postcode={230026},
    country={China}}

\affiliation[pku]{organization={School of Physics, Peking University},city={Beijing},postcode={100871},country={China}}


\affiliation[USTCdep]{
    organization={Department of Modern Physics, University of Science and Technology of China},
    city={Hefei},
    postcode={230026},
    country={China}}



\affiliation[YaLongSD]{
    organization={Yalong River Hydropower Development Company},
    city={Chengdu},
    postcode={610051},
    country={China}}

\affiliation[IAP]{
    organization={Shanghai Institute of Applied Physics, Chinese Academy of Sciences},
    city={Shanghai},
    postcode={201800},
    country={China}}

\affiliation[CHEPpku]{
    organization={Center for High Energy Physics, Peking University},
    city={Beijing},
    postcode={100871},
    country={China}}

\affiliation[SDUdep]{
    organization={Research Center for Particle Science and Technology, Institute of Frontier and Interdisciplinary Science, Shandong University},
    city={Qingdao},
    postcode={266237},
    country={China}}

\affiliation[SDUlab]{
    organization={Key Laboratory of Particle Physics and Particle Irradiation of Ministry of Education, Shandong University},
    city={Qingdao},
    postcode={266237},
    country={China}}

\affiliation[UMD]{
    organization={Department of Physics, University of Maryland, College Park},
    city={Maryland},
    postcode={20742},
    country={USA}}

\affiliation[MESJTU]{
    organization={School of Mechanical Engineering, Shanghai Jiao Tong University},
    city={Shanghai},
    postcode={200240},
    country={China}}

\affiliation[SYUSFI]{
    organization={Sino-French Institute of Nuclear Engineering and Technology, Sun Yat-sen University},
    city={Zhuhai},
    postcode={519082},
    country={China}}

\affiliation[NKU]{
    organization={School of Physics, Nankai University},
    city={Tianjin},
    postcode={300071},
    country={China}}

\affiliation[YTU]{
    organization={Department of Physics, Yantai University},
    city={Yantai},
    postcode={264005},
    country={China}}

\affiliation[FDU]{
    organization={Key Laboratory of Nuclear Physics and Ion-beam Application (MOE), Institute of Modern Physics, Fudan University},
    city={Shanghai},
    postcode={200433},
    country={China}}

\affiliation[USST]{
    organization={School of Medical Instrument and Food Engineering, University of Shanghai for Science and Technology},
    city={Shanghai},
    postcode={200093},
    country={China}}

\affiliation[SPEIT]{
    organization={SJTU Paris Elite Institute of Technology, Shanghai Jiao Tong University},
    city={Shanghai},
    postcode={200240},
    country={China}}

\affiliation[NNU]{
    organization={School of Physics and Technology, Nanjing Normal University},
    city={Nanjing},
    postcode={210023},
    country={China}}

\affiliation[SYSUzhuhai]{
    organization={School of Physics and Astronomy, Sun Yat-sen University},
    city={Zhuhai},
    postcode={519082},
    country={China}}

\affiliation[CDUT]{
    organization={College of Nuclear Technology and Automation Engineering, Chengdu University of Technology},
    city={Chengdu},
    postcode={610059},
    country={China}}

\author[SYU]{Shu Zhang}
\author[shKeyLab]{Zihao Bo}
\author[shKeyLab]{Wei Chen}
\author[TDLee,shKeyLab,SJTUSC,scKeyLab]{Xun Chen}
\author[YaLongSD,scKeyLab]{Yunhua Chen}
\author[SYUSFI]{Zhaokan Cheng}
\author[TDLee]{Xiangyi Cui}
\author[YTU]{Yingjie Fan}
\author[FDU]{Deqing Fang}
\author[shKeyLab]{Zhixing Gao}
% ...

\cortext[cor1]{Corresponding author}
\cortext[cor2]{Spokesperson}

\begin{abstract}
We report the search for neutrinoless double-beta decay of $^{136}$Xe from the PandaX-4T experiment with a 3.7-tonne natural xenon target.
The data reconstruction and the background modeling are optimized in the MeV energy region. 
A blind analysis is performed with data from the commissioning run and the first science run. 
No significant excess of signal over the background is observed. 
A lower limit on the half-life of $^{136}$Xe neutrinoless double-beta decay is established to be $2.1 \times 10^{24}$~yr at the 90\% confidence level, with a $^{136}$Xe exposure of 44.6~kg$\cdot$year. 
Our result represents the most stringent constraint from a natural xenon detector to date.
\end{abstract}


\end{frontmatter}

\section{Introduction}
The two-neutrino double-beta decay ($2\nu \beta \beta$) in a nucleus is a second-order weak transition allowed by the Standard Model (SM) of particle physics, in which two neutrons decay simultaneously into two protons, emitting two kinetic electrons and two antineutrinos~\cite{goeppert1935double}.
This process has been observed in 11 nuclides so far~\cite{barabash2015average}.
On the other hand, the neutrinoless double-beta decay ($0\nu \beta \beta$), proposed by Furry in 1937~\cite{furry1939transition}, would only have two electrons emitted with a total kinetic energy equal to the Q-value of the reaction (omitting the minuscule nuclear recoil energy), without accompanying antineutrinos.
The observation of $0\nu \beta \beta$ would reveal the Majorana nature of neutrinos, which could offer an explanation of the matter-dominance in the Universe~\cite{Fukugita:1986hr}.
The effective Majorana mass of neutrinos is linked to the neutrino mass ordering as well as the absolute neutrino masses~\cite{dolinski2019neutrinoless,cremonesi2014challenges,avignone2008double}.
$0\nu \beta \beta$ is also a direct evidence for physics beyond the SM that violates lepton number conservation~\cite{dolinski2019neutrinoless}.
Numerous experiments have searched for $0\nu \beta \beta$ in different isotopes, for instance $^{76}$Ge~\cite{agostini2020final,arnquist2023final,CDEX:2023owy}, $^{130}$Te~\cite{CUORE:2021mvw}, and $^{136}$Xe~\cite{anton2019search,abe2024search}.
To date, the most stringent lower limits on $0\nu \beta \beta$ half-lives are on the order of $10^{26}$ years.

Dual-phase xenon time projection chambers (TPCs) have been successfully used in the dark matter direct detection experiments, in which the main focus is on the keV region~\cite{PandaX:2024qfu,XENON:2023cxc,LZCollaboration:2024lux}.
With an 8.9\% isotopic abundance of $^{136}$Xe, the current multi-tonne natural xenon experiments are reaching the similar $^{136}$Xe target mass as those of $^{136}$Xe-enriched $0\nu \beta \beta$ experiments.
Searches for $^{136}$Xe $0\nu \beta \beta$ with natural xenon have been performed in PandaX-II~\cite{ni2019searching} and XENON1T~\cite{aprile2022double}. 
Previously, we have precisely measured the $2\nu \beta \beta$ half-life of $^{136}$Xe based on the PandaX-4T commissioning run (Run0) data~\cite{si2022determination}. 
In this paper, we report new results of searching for $^{136}$Xe $0\nu\beta\beta$ from a blind analysis of PandaX-4T Run0 and the first science run (Run1) data with a total $^{136}$Xe exposure of 44.6 kg$\cdot$year, based on a unified data processing pipeline with an improved background modeling.

\section{PandaX-4T Detector}
The PandaX-4T detector is a dual-phase Xe TPC located in the Hall B2 of the China Jinping Underground Laboratory (CJPL)~\cite{jainmin2015second}. 
The cylindrical TPC contains 3.7~tonnes of liquid xenon (LXe) in the active volume.
The volume is surrounded by 24 Polytetrafluoroethylene (PTFE) reflective panels on the side with a diameter of 1185~mm.
A gate electrode on the top and a cathode on the bottom are separated by 1168~mm. 
An anode is positioned 10~mm above the gate electrode, with the liquid xenon level maintained in between.
Two photomultiplier tube (PMT) arrays are placed on the top and bottom of the TPC, equipped with 169 and 199 three-inch Hamamatsu R11410-23 PMTs, respectively. 
The LXe detector is enclosed in a dual-layer cryostat, which is supported by a stainless steel platform (SSP).
The setup is immersed in the center of a cylindrical ultrapure water shielding tank, which is 13~m high and 10~m in diameter.
Physical events in the active volume generate prompt scintillation light ($S1$) and ionized electrons; the latter drift upwards in an electric field and are then extracted into the gaseous xenon region under a much stronger electric field, where the electrons undergo an electroluminescence process, producing delayed scintillation light ($S2$). 
The time difference of $S2$ and $S1$ measures the vertical position (Z) of an event, and the $S2$ hit pattern of the top PMTs is used to reconstruct the horizontal position (X-Y) of the event.
More details about the PandaX-4T detector and the operation parameters can be found in Ref.~\cite{PandaX-4T:2021bab,PandaX:2024qfu}.

\section{Data Reconstruction and Event Selection}
The analysis pipeline in the MeV energy region is based on the previous works for Run0~\cite{si2022determination,yan2023searching}.
A $S2$ waveform slicing algorithm is developed to identify multiple $S2$s overlapping in time, resulting in better discrimination of single-site (SS) and multiple-site (MS) events.
PMT saturation is corrected by matching the rising slope of the saturated waveforms to those of the non-saturated waveforms in the same event~\cite{Luo:2023ebw}.
The X-Y positions of events are reconstructed based on the hit pattern of desaturated $S2$ from the top PMTs, using the photon acceptance function (PAF) method with multiple layers of image PMTs to account for the PTFE reflection~\cite{zhang2021horizontal}.
The position-dependent detection efficiency of $S1$ is corrected in three dimensions based on the uniformly distributed $^{83m}$Kr calibration events. 
The same method is applied to $S2$ correction in the X-Y plane.
The magnitude of the $S2$ Z-dependency is corrected by the drifting electron lifetime, which is calculated from the uniformly distributed 5.6~MeV $\alpha$ events of $^{222}$Rn~\cite{PandaX:2024med}.
The temporal correction is applied separately to $S1$ and $S2$ on a daily basis, and the correction factors are also determined from $^{222}$Rn events~\cite{PandaX:2024med}.
Energy is calibrated with internal and external radioactive sources and reconstructed as $E = 13.7~\mathrm{eV} \times (S1/\mathrm{PDE}+S2_\mathrm{B}/(\mathrm{EEE}\times \mathrm{SEG_B}))$~\cite{szydagis2011nest}, where $S2_\mathrm{B}$ stands for $S2$ detected by the bottom PMTs, and PDE, EEE, and $\mathrm{SEG_B}$ represent the $S1$ photon detection efficiency, the $S2$ electron extraction efficiency, and the single-electron gain for $S2_\mathrm{B}$, respectively. 
Residual energy nonlinearity is corrected by an empirical third-order polynomial based on characteristic $\gamma$ peaks.

The analysis pipeline is further optimized in two major aspects.
Firstly, approximately 0.5\% of SS events have been successfully recovered with an improved time window cut.
When two SS events occur closely in time, there is a risk that they may be incorrectly recognized as an MS event, leading to a loss of data. 
We have developed an algorithm based on the MC to effectively separate such events while ensuring the appropriate time window to preserve genuine MS events. 

Secondly, a 3.5~ms dead-time window before any $^{214}$Po event is enforced to remove $^{214}$Bi events.
Due to the improved event time window cut, $^{214}$Bi-$^{214}$Po cascade decays are typically separated in two consecutive events, resulting in more isolated $^{214}$Bi SS events.
Similar to the $S2$ waveform slicing method~\cite{si2022determination}, we have developed a $S1$ waveform slicing algorithm for $\alpha$ events, which usually have long tails compared to electronic recoil (ER) events and could be wrongly tagged.
After the $S1$ waveform slicing, $^{214}$Po events are identified according to the charge and number of $S1$, with an efficiency of 99.7\%.
All ER events within the dead-time window before $^{214}$Po events are removed, leading to a background reduction of about 1\% from the veto of isolated $^{214}$Bi. 
The corresponding live-time loss is negligible due to the low $^{214}$Po rate of about 1~mHz.

Run0 and Run1 data are processed with the optimized data production pipeline with reconstruction and correction parameters determined from the calibration data.
We blind the $^{136}$Xe $0\nu\beta\beta$ signal region of interest (ROI) from 2356~keV to 2560~keV defined as $Q_{\beta\beta}^{^{136}Xe}\pm2\sigma$, where $Q_{\beta\beta}^{^{136}Xe}$ is the Q-value and $\sigma$ is the absolute energy resolution.
The energy range for spectrum fit is from 1100~keV to 2800~keV.
The low-energy boundary is set to avoid the contribution of $^{214}$Pb, a progeny of $^{222}$Rn emanating from materials of the detector and auxiliary subsystems, which has a $\beta$ decay end-point of 1018~keV.
Above 2800~keV, the uncertainty of the SS fraction defined as the ratio of SS events to the total events is too large due to limited statistics in the calibration data.
Quality cuts to remove nonphysical events and select ER events are adopted from Ref.~\cite{si2022determination} and verified in the newly processed Run0 and Run1 data.
The cut efficiencies in the energy range of spectrum fit are estimated as $(99.89\pm0.10)\%$ and $(99.97\pm0.02)\%$ for Run0 and Run1 data, respectively.

\begin{figure}[tbp]
    \centering
    \includegraphics[width=\linewidth]{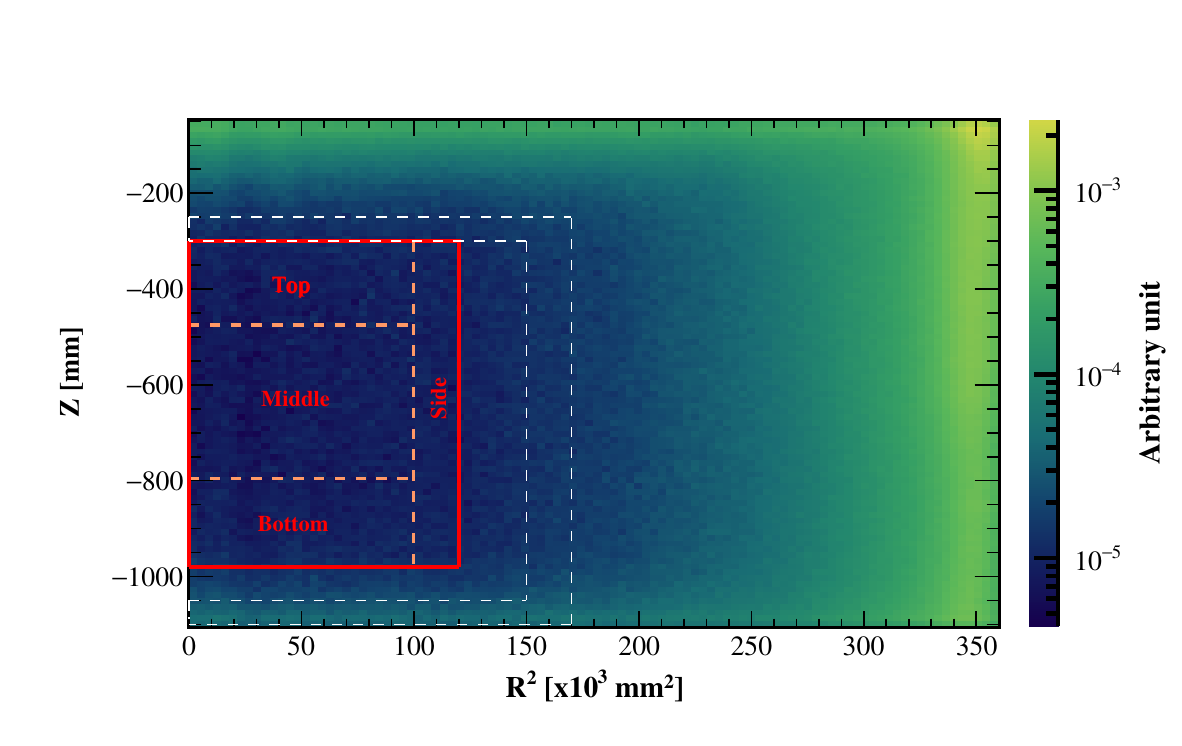}
    \caption{Distribution of the positions of reconstructed physics events. The solid red lines represent the boundary of the optimized FV for both Run0 and Run1, with dashed orange lines dividing the FV into 4 regions. The dashed white lines define the outer region, in which the detector response is modeled.}
    \label{fig:FV}
\end{figure}

The fiducial volume (FV) is defined by maximizing the figure-of-merit $FoM = m/\sqrt{B}$ in the ROI, where $m$ is the LXe mass of FV, proportional to the counts of $^{136}$Xe $0\nu \beta \beta$ signal, and $B$ is the counts of background events, estimated from the background model which will be discussed later.
The boundary of the optimized FV is shown as the solid red lines in Fig.~\ref{fig:FV} for both Run0 and Run1.
The uncertainty of FV due to the limited position reconstruction precision is dominated by the uncertainty in R directions.
Calculated from the non-uniformity of the $^{83m}$Kr calibration data distribution in the R direction~\cite{yan2023searching,PandaX:2024qfu}, an uncertainty of $0.4\%$ for Run0 and $1.9\%$ for Run1 is observed.
Additional FV uncertainty is due to the LXe density fluctuation, which is estimated as $(2.850\pm0.004)$~g/cm$^{3}$ based on the detector operating pressure~\cite{si2022determination}.
The final FV mass is $(735\pm3)$~kg for Run0 and $(735\pm14)$~kg for Run1.
Similar to Ref.~\cite{si2022determination}, the FV is further divided into four regions, with each containing nearly the same number of events in the spectrum fitting range, in order to better constrain the background from the top, bottom, and side of the detector.

Detector energy response is modeled and used in the final spectrum fit.
To account for the potential shift between the reconstructed energy $E$ and the simulated energy $\hat{E}$, the energy scale is defined as 
\begin{equation}
 E  = a\cdot \hat{E}^2+b\cdot \hat{E}+c. \label{func:deltaE}
\end{equation}
Then, following the same procedure as that in Ref.~\cite{yan2023searching}, the energy resolution $\sigma$ as a function of the reconstructed energy $E$ is modeled as
\begin{equation}
    \frac{\sigma(E)}{E}  = \frac{d}{\sqrt{E}}+e\cdot E+f.
 \label{func:sigmaE} 
\end{equation}

The parameters $\mathcal{M}=(a,b,c,d,e,f)^T$ are obtained from fitting the characteristic $\gamma$ peaks of 164~keV ($^{131m}$Xe), 236~keV ($^{127}$Xe and $^{129m}$Xe), 1460~keV ($^{40}$K), and 2615~keV ($^{208}$Tl) of an outer region shown in Fig.~\ref{fig:FV}.
The nominal values $\mathcal{M}_0$ and the covariance matrix $\Sigma_\mathcal{M}$ are implemented in the likelihood function for spectrum fitting.
Table.~\ref{table:sys_err} shows the nominal values of each parameter and their respective uncertainties.

In addition to the energy response model, systematic uncertainties are also shown in the Table.~\ref{table:sys_err}, originated from the overall efficiency and the $^{136}$Xe mass.
The overall efficiency is calculated as the product of quality cut efficiency and SS fraction (see details later) with the uncertainty propagated quadratically.
The $^{136}$Xe mass is evaluated from the FV mass and the isotopic abundance. 
The latter is measured to be $(8.58 \pm 0.11)\%$ by a residual gas analyzer with xenon samples taken from the detector.

\begin{table}[tbp]
    \centering
    \caption{Summary of systematic uncertainties from the energy response model, the overall efficiency, and the $^{136}$Xe mass. 
 Due to the lower statistics of characteristic $\gamma$ peaks, the uncertainties of the energy response model in Run1 are larger than those in Run0. 
 The uncertainties of the background model in the ROI are listed in detail in Table.~\ref{table:fitCounts}.}
    \renewcommand{\arraystretch}{1.5}
    \resizebox{\linewidth}{!}{
   \begin{tabular}{>{\centering\arraybackslash}m{1.cm}>{\centering\arraybackslash}m{1.cm}>{\centering\arraybackslash}m{1.cm}>{\centering\arraybackslash}m{1.cm}}
       \hline
       \hline
       \multicolumn{2}{c}{\multirow{2}{*}{Sources}} & \multicolumn{2}{c}{\multirow{1}{*}{Values}} \\ 
       \cline{3-4}
       \multicolumn{2}{c}{\multirow{1}{*}{}} & \multicolumn{1}{c}{\multirow{1}{*}{Run0}} & \multicolumn{1}{c}{\multirow{1}{*}{Run1}}\\
       \hline
       \multicolumn{1}{c}{\multirow{6}{*}{\shortstack[t]Energy response}}&
       \multicolumn{1}{c}{\multirow{1}{*}{\shortstack[t]{$a$~[keV$^{-1}$]}}} & \multicolumn{1}{c}{$(4.2\pm1.0) \times 10^{-6}$}& \multicolumn{1}{c}{$(1.1\pm1.4) \times 10^{-6}$} \\
       \multicolumn{1}{c}{\multirow{1}{*}{}}&\multicolumn{1}{c}{\multirow{1}{*}{\shortstack[t]{$b$}}} & \multicolumn{1}{c}{$0.992\pm0.002$}& \multicolumn{1}{c}{$0.997\pm0.004$} \\
       \multicolumn{1}{c}{\multirow{1}{*}{}}&\multicolumn{1}{c}{\multirow{1}{*}{\shortstack[t]{$c$~[keV]}}} & \multicolumn{1}{c}{$0.90\pm0.32$}& \multicolumn{1}{c}{$1.4\pm1.5$} \\
       \cline{2-4}
       \multicolumn{1}{c}{\multirow{1}{*}{}}&
       \multicolumn{1}{c}{\multirow{1}{*}{\shortstack[t]{$d$~[$\sqrt{\mathrm{keV}}$]}}} & \multicolumn{1}{c}{$0.259\pm0.046$} & \multicolumn{1}{c}{$0.46\pm0.25$} \\
       \multicolumn{1}{c}{\multirow{1}{*}{}}&\multicolumn{1}{c}{\multirow{1}{*}{\shortstack[t]{$e$~[keV$^{-1}$]}}} & \multicolumn{1}{c}{$(1.1\pm1.5) \times 10^{-6}$} & \multicolumn{1}{c}{$(8.8\pm22.2) \times 10^{-7}$} \\
       \multicolumn{1}{c}{\multirow{1}{*}{}}&\multicolumn{1}{c}{\multirow{1}{*}{\shortstack[t]{$f$}}} & \multicolumn{1}{c}{$(9.7\pm3.5) \times 10^{-3}$} & \multicolumn{1}{c}{$(7.4\pm10.0) \times 10^{-3}$} \\
       \cline{1-4}
       \multicolumn{1}{c}{\multirow{2}{*}{\shortstack[t]{Overall efficiency}}}&\multicolumn{1}{c}{$^{136}$Xe 0$\nu\beta\beta$ SS fraction} & \multicolumn{1}{c}{($87.1\pm11.3$)\%}& \multicolumn{1}{c}{($87.3\pm7.0$)\%} \\
       \multicolumn{1}{c}{\multirow{2}{*}{}}&\multicolumn{1}{c}{Quality cut} & \multicolumn{1}{c}{($99.89\pm0.10$)\%}& \multicolumn{1}{c}{($99.97\pm0.02$)\%} \\
       \cline{1-4}
       \multicolumn{1}{c}{\multirow{2}{*}{\shortstack[t]{$^{136}$Xe mass}}}&\multicolumn{1}{c}{$^{136}$Xe abundance} & \multicolumn{2}{c}{($8.58\pm0.11$)\%} \\
       \multicolumn{1}{c}{\multirow{2}{*}{}}&\multicolumn{1}{c}{FV mass~[kg]} & \multicolumn{1}{c}{$735\pm3$}& \multicolumn{1}{c}{$735\pm14$} \\
       \cline{1-4}
       \multicolumn{2}{c}{Background model} & \multicolumn{2}{c}{Table.~\ref{table:fitCounts}} \\
       \hline
       \hline
   \end{tabular}
 }
   \label{table:sys_err}
\end{table}

\section{Signal and Background Models}
\begin{figure}[tb]
    \centering
        \includegraphics[width=\linewidth]{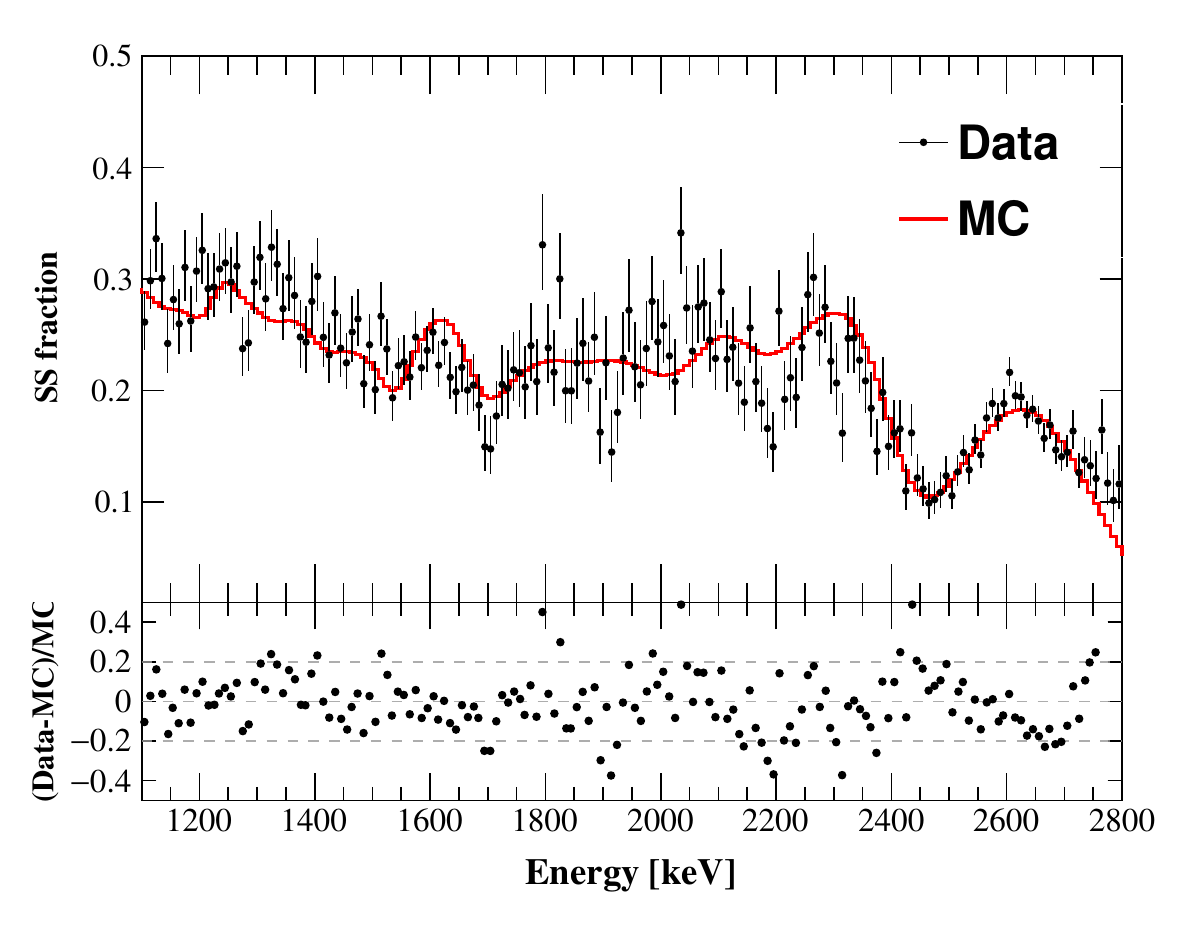}
        \vspace{3pt}
        \includegraphics[width=\linewidth]{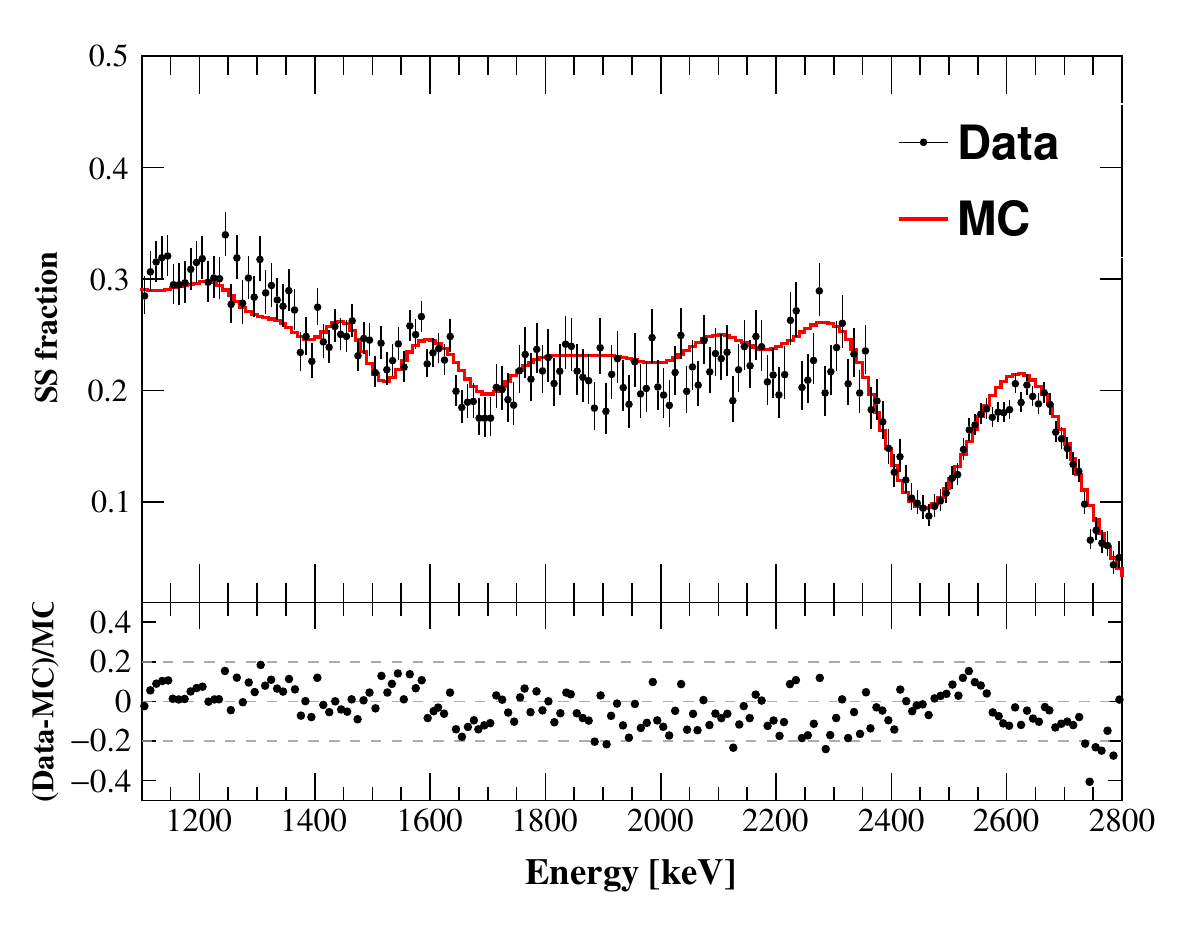}
    \caption{Comparison of SS fractions between simulation and data of $^{232}$Th calibration for Run0 (top) and Run1 (bottom) with relative differences between mean values shown in the lower panels. The spectrum-averaged deviations in absolute values are conservatively taken as relative uncertainties for the SS fractions, which are 13\% for Run0 and 8\% for Run1.}
    \label{fig:ssErr}
\end{figure}

The $^{136}$Xe $0\nu\beta\beta$ signals are generated from the Decay0 package~\cite{ponkratenko2000event} and then fed to BambooMC~\cite{chen2021bamboomc}, a custom-designed Monte Carlo simulation framework based on Geant4.
The detector energy response is modeled in the signal generation with $\mathcal{M}$. 
To distinguish the SS and MS events of the signals, a $S2$ waveform simulation algorithm is adopted from a previous study~\cite{si2022determination} to generate pseudo-$S2$ waveforms with parameters from data production.
The simulated $S2$ waveforms are then processed through the same data production pipeline to select SS signal events.
The resulting SS signal fraction is 87.1\% for Run0 and 87.3\% for Run1.
The uncertainty of the SS fraction is evaluated by comparing the simulation and data of the $^{232}$Th calibration campaign.
Fig.~\ref{fig:ssErr} shows the bin-by-bin comparison of SS fractions between simulation and calibration data, and the spectrum-averaged deviation (absolute value) of their mean values is conservatively taken as the relative uncertainty, which is $13\%$ for Run0 and $8\%$ for Run1. 
The signal SS fractions within the ROI with the corresponding uncertainties are listed in Table.~\ref{table:sys_err}.
We only utilize the SS spectrum for the final spectrum fit.

The background in the energy range of 1100 to 2800~keV mainly comes from three sources, including $^{136}$Xe $2\nu\beta\beta$, the detector material, and the SSP.
The half-life of $^{136}$Xe $2\nu\beta\beta$ has been measured in a previous study~\cite{si2022determination}, while the energy spectrum of two electrons is generated from Decay0.
The radioactivity of $^{60}$Co, $^{40}$K, $^{232}$Th, and $^{238}$U from the detector material have been measured by high-purity germanium (HPGe) counting stations~\cite{qian2022low}.
The detector is grouped as top, bottom, and side parts~\cite{si2022determination}.
Although partially shielded by the ultrapure water, the SSP with a total weight of more than 6.8~tonnes still contributes a significant background above 2~MeV, as discussed in Ref.~\cite{dbdes}.
The $^{232}$Th and $^{238}$U radioactivity of SSP is measured \emph{in situ} by an MS energy spectrum fit~\cite{dbdes} instead of HPGe since the SSP consists of raw materials from various sources and only a fraction was counted by HPGe. 
The shape of $^{60}$Co and $^{40}$K from the SSP are highly degenerate with those from the detector material in the spectrum fit range and thus excluded in the background model. 
The contribution of all the background sources is modeled with BambooMC and processed with the data production pipeline. 
The uncertainties are included as Gaussian penalties to constrain the corresponding nuisance parameters in the likelihood fit.

Other background contributions are also considered.
As noted earlier, residual $^{214}$Bi background has been suppressed by the specific data selection cut.
Including $^{214}$Bi $\beta$ and $\gamma$ contributions in the SS spectrum fitting to the blinded data shows negligible impact.
The $\gamma$'s from $^{214}$Bi originating from LXe outside the TPC active volume might enter the FV and contribute to the background. 
However, this contribution is negligible, based on the simulation and the measured $^{214}$Po activity in the LXe TPC.
The 2505~keV peak from $^{60}$Co is especially important since it resides in the ROI.
The 1173~keV and 1332~keV cascade $\gamma$'s from $^{60}$Co would be reconstructed as a 2505~keV SS event if the energy depositions occur in nearly the same Z positions.
Comparison between $^{60}$Co calibration data and simulation shows a consistent ratio of 2505~keV and 1332~keV peaks, confirming a reliable estimation of this background.

The background within the ROI is dominated by the radioactivity of the detector material and the SSP.
The most significant contribution is from the left slope of the full absorption peak and its Compton scattering of 2615~keV $\gamma$ from $^{208}$Tl in the $^{232}$Th chain.
The full absorption peak of 2448~keV $\gamma$ from $^{214}$Bi in the $^{238}$U chain is also noticeable.
There is additional background from the $^{60}$Co 2505~keV peak.
The expected background counts within the ROI are shown in Table.~\ref{table:fitCounts}.
In the table, we also listed the contribution from $^{136}$Xe $2\nu\beta\beta$, which is negligible in the ROI.

\begin{table}[tbp]
    \centering
    \caption{Comparison of background counts in the ROI from the expectation of the background model, expectation of the blinded best-fit, and the unblinded best-fit results.}
    \renewcommand{\arraystretch}{1.5}
    \centering
    \resizebox{\linewidth}{!}{
    \begin{tabular}{c|ccc}
    \hline
    \hline
 Background & Model expectation & Blinded fit & Unblinded fit \\ \hline
 SSP $^{232}$Th& $527\pm45$& $470\pm34$& $458\pm33$\\
 SSP $^{238}$U& $50\pm15$& $38\pm11$&$39\pm11$\\ 
    $^{232}$Th & $375\pm224$& $510\pm34$& $485\pm31$\\
 $^{238}$U & $78\pm42$& $70\pm9$&$72\pm9$\\ 
    $^{60}$Co & $18\pm7$ & $31\pm3$ & $31\pm3$ \\ 
    $^{136}$Xe & $0.18\pm0.01$ & $0.19\pm0.01$ & $0.19\pm0.01$ \\ 
    \hline \hline
    \end{tabular}
 }
    \label{table:fitCounts}
\end{table}

\section{Fitting Method and Results}
We employ a binned Poisson likelihood method for spectra fitting.
The likelihood function is constructed as follows:
\begin{equation}
    \begin{aligned} \hspace{-2em}
 L =& \prod_{r}^{N_{run}}\prod_{i}^{N_{region}} \prod_{j}^{N_{bins}} \frac{(N_{rij})^{N_{rij}^{obs}}}{N_{rij}^{obs}!}e^{-N_{rij}} \\
 &  \cdot\prod_{r}^{N_{run}} [\mathcal{G}(\mathcal{M}_{r}; \mathcal{M}_{r}^{0},\Sigma_{r}^{\mathcal{M}}) \cdot \prod_{k}^{N_{eff}} G(\eta_r^{k};0, \sigma_r^{k})] \\
 & \cdot \prod_{b}^{N_{bkg}} G(\eta^b;0, \sigma^b)    
    \end{aligned}    
\end{equation}
where \(N_{rij}\) and \(N_{rij}^{obs}\) represent the expected and observed number of events in the $j$-th bin of the $i$-th region in the $r$-th Run (Run0 or Run1), respectively. 
\(N_{rij}\) is defined as:
\begin{equation}
    \begin{aligned}
 N_{rij} =& (1+\eta_r^{o})\cdot [(1+\eta_{r}^{s}) \cdot n_{r}^{s} \cdot S_{ijr} \\
 &+ \sum_{b}^{N_{bkg}} (1+\eta^b) \cdot n_{r}^{b} \cdot B_{ijr}^{b}]
    \end{aligned}
\end{equation}
where $n_{r}^{s}$ and $n_{r}^{b}$ are the counts of $^{136}$Xe $0\nu\beta\beta$ signal and the background component $b$, respectively.
$S_{ijr}$ and $B_{ijr}^{b}$ represent the normalized energy spectra for the signal and background component $b$, respectively, both are convolved with the energy response model of six-parameter $\mathcal{M}_{r}$.
$\eta^{b}$ is the nuisance parameter of $n_{r}^{b}$, which represents the fractional deviation between the fitted value and the expected value.  
The nuisance parameter $\eta_r^{o}$ and $\eta_{r}^{s}$ represent the systematic uncertainties of the overall efficiency and the $^{136}$Xe mass in the $r$-th Run, respectively.
All nuisance parameters are constrained individually by Gaussian penalty terms $G(\eta_r^{k};0, \sigma_r^{k})$, except that $\mathcal{M}_{r}$ is correlatively constrained by the six-dimensional $\mathcal{G}(\mathcal{M}_{r}; \mathcal{M}_{r}^{0},\Sigma_{r}^{\mathcal{M}})$ with the nominal values $\mathcal{M}_{r}^0$ and the covariance matrix $\Sigma_{r}^\mathcal{M}$ in the $r$-th Run.

\begin{figure}[tb]
    \centering
        \includegraphics[width=\linewidth]{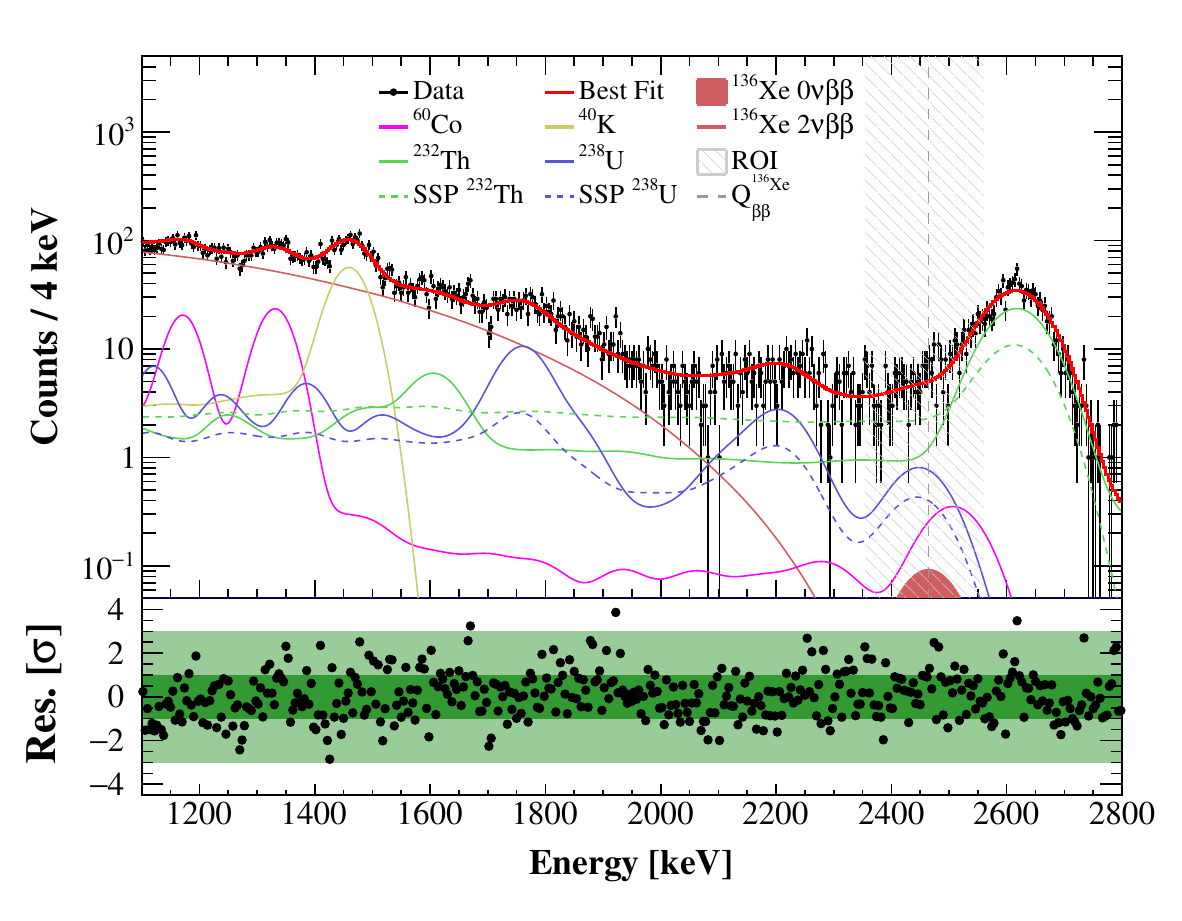}
        \includegraphics[width=\linewidth]{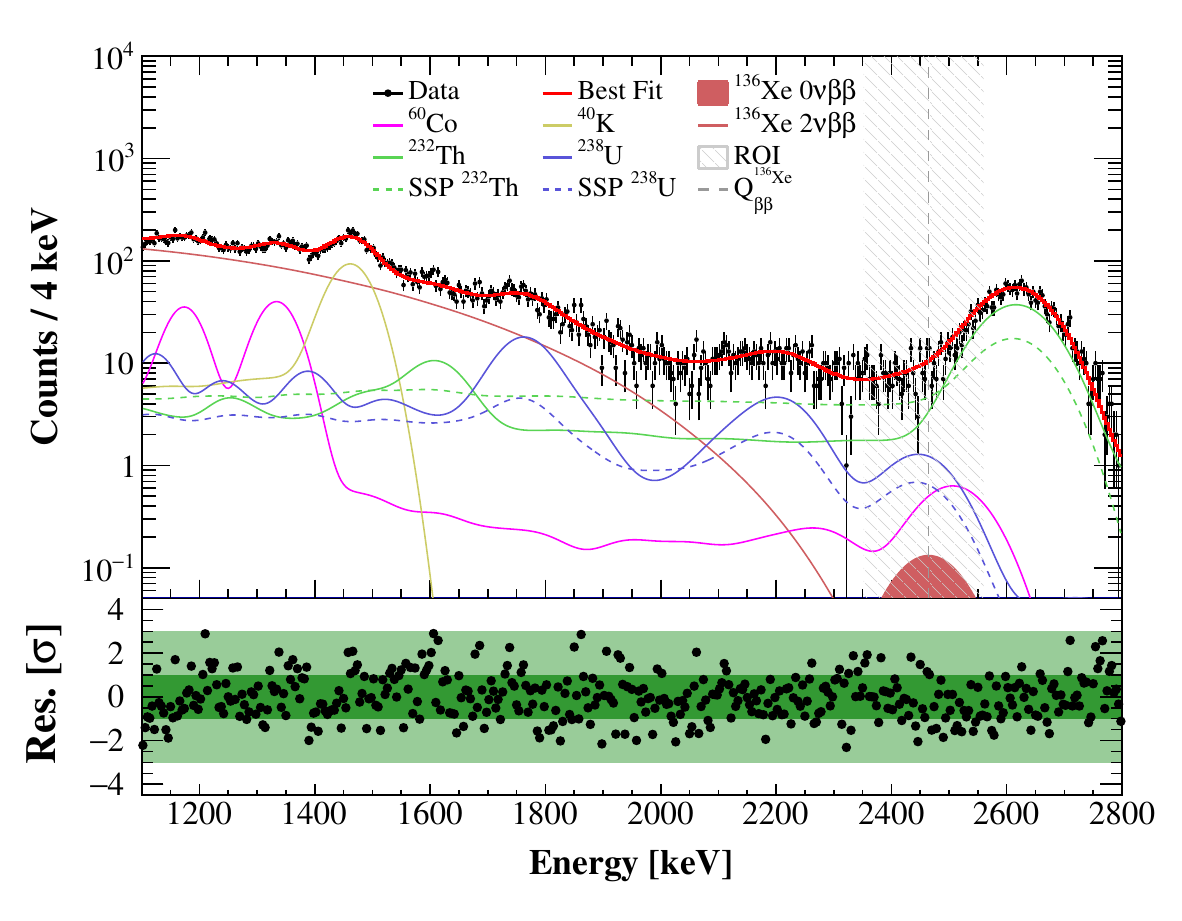}
    \caption{Unblinded fit results for Run0 (top) and Run1 (bottom). The contributions from $^{60}$Co, $^{40}$K, $^{232}$Th, and $^{238}$U of different detector parts are combined respectively after fitting for better visualization. $^{136}$Xe $0\nu\beta\beta$ signals are indicated by red-shaded peaks. The lower panel shows the residuals together with the $\pm$1$\sigma$ ($\pm$3$\sigma$) region represented by the dark (light) green band.}
    \label{fig:unblindedFit}
\end{figure}
\begin{figure}[tb]
    \centering
    \includegraphics[width=0.9\linewidth]{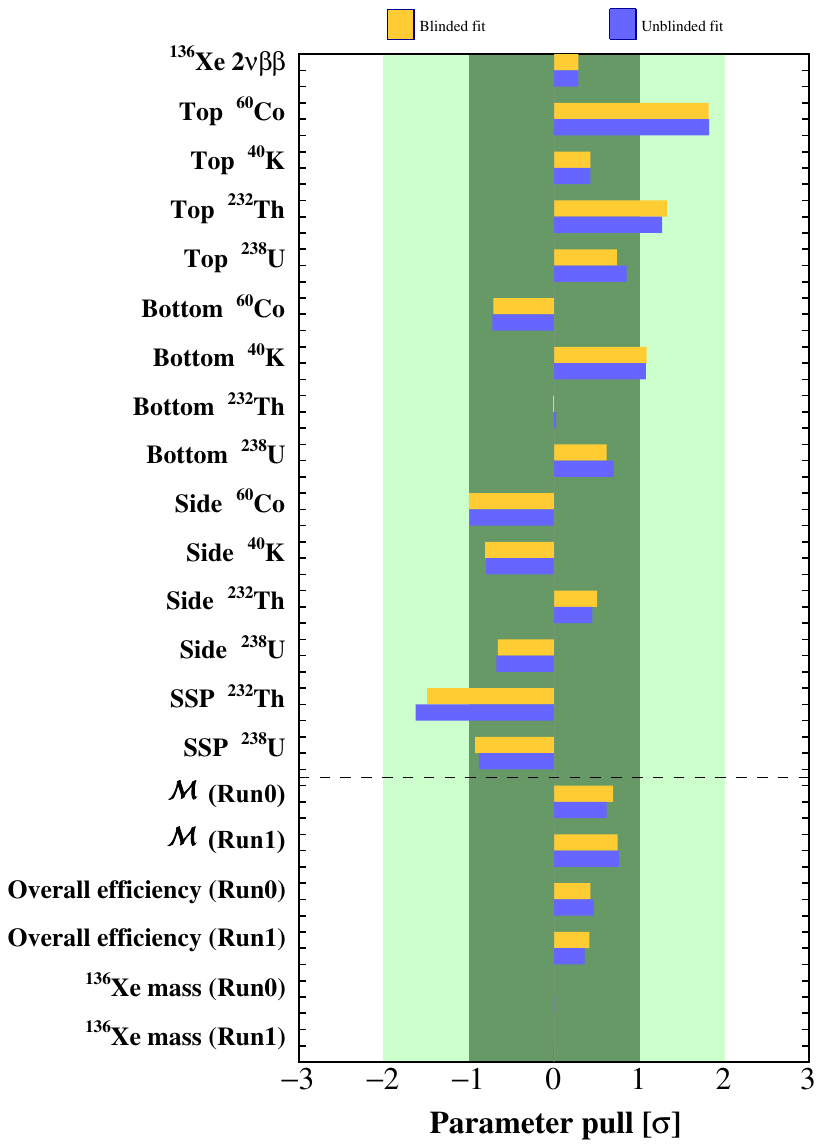}
    \caption{Pulls of nuisance parameters in the unit of $\sigma$ for blinded fits (yellow bar) and unblinded fits (blue bar). The nuisance parameters of $^{136}$Xe mass only appear in the unblinded fits. The dark (light) green band represents the $\pm1\sigma$ ($\pm2\sigma$) region. All pulls fall within the $\pm2\sigma$ region, and no significant discrepancy arises between blinded and unblinded fit.}
    \label{fig:pull}
\end{figure}

Prior to unblinding, we performed a background-only fit to the spectra, excluding the ROI.
The goodness-of-fit is $\chi^{2}/\text{ndf}=1.14$.
The expected counts of background within the ROI are extrapolated and reported in Table.~\ref{table:fitCounts}.
Toy data are generated from the best-fit results and then fitted with both signal and background models.
The median sensitivity of the lower limit on $^{136}$Xe $0\nu\beta\beta$ half-life is estimated as $2.7 \times 10^{24}$~yr at the $90\%$ confidence level (C.L.).

A spectral fit with the signal plus background models is then performed after we unblind data in the ROI, as shown in Fig.~\ref{fig:unblindedFit}.
Corresponding fitting results of the four regions in FV are shown in~\ref{appen}.
The goodness-of-fit value for this case is $\chi^{2}/\text{ndf}=1.15$.
The best-fit $^{136}$Xe $0\nu\beta\beta$\ decay rate is $(14\pm37)$~t$^{-1}$yr$^{-1}$, with a $p$ value of 0.49 for null result.
Therefore, the lower limit on the $^{136}$Xe $0\nu\beta\beta$ half-life is derived as $T^{0\nu\beta\beta}_{1/2} > 2.1 \times 10^{24}$~yr at the $90\%$ C.L., consistent with the median sensitivity within 1.1$\sigma$.
The corresponding upper limit on effective Majorana neutrino mass is in the range of $\langle m_{\beta\beta} \rangle = (0.4-1.6)$~eV/c$^{2}$, taken into account the phase space factor calculation~\cite{Stoica:2013lka} and various nuclear matrix element (NME) calculations~\cite{Pompa:2023jxc}. 

The pulls of nuisance parameters fall within the $\pm 2\sigma$ range for both the blinded and unblinded fits, as illustrated in Fig.~\ref{fig:pull}.
All best-fit nuisance parameters are consistent between the blinded and unblinded fits, confirming the stability of the fitting process as well as the proper modeling of background and energy response.
Note that the pull of $^{60}$Co from the top part of the detector material reaches 1.8$\sigma$, indicating that this background component estimated from the HPGe material assay might be slightly underestimated.
Fitted background counts combined from Run0 and Run1 in the ROI are also reported in Table.~\ref{table:fitCounts}.
The energy resolution at 2615~keV in FV is fitted to be 2.0\% for Run0 and 2.3\% for Run1.

\section{Summary}
In summary, we present a blind analysis to search for $^{136}$Xe $0\nu\beta\beta$ with a combined dataset of PandaX-4T Run0 and Run1.
A unified analysis pipeline for Run0 and Run1 is established in the MeV region with optimized data reconstruction and event selection.
The energy response of the detector is modeled and implemented analytically into the likelihood function.
The background contributions are modeled from prior measurements and determined \emph{in situ} by spectral fitting. 

No significant excess of signal over the background is observed after unblinding the data.
The lower limit on $^{136}$Xe $0\nu\beta\beta$ half-life is achieved as $T^{0\nu\beta\beta}_{1/2} > 2.1 \times 10^{24}$~yr at the $90\%$ C.L., corresponding to the upper limit on effective Majorana neutrino mass in the range of $\langle m_{\beta\beta} \rangle = (0.4-1.6)$~eV/c$^{2}$.
Our results highlight an improvement to our previous PandaX-II results~\cite{ni2019searching} by an order of magnitude and to the XENON1T results~\cite{aprile2022double} by a factor of 1.8.
This represents so far the most stringent constraint of $^{136}$Xe $0\nu\beta\beta$ half-life from natural xenon detectors, demonstrating the potential of $^{136}$Xe $0\nu\beta\beta$ search with next-generation multi-ten-tonne natural xenon detectors.

After a long shutdown, PandaX-4T has resumed physics data-taking with an upgraded detector.
More statistics of data with improved detector performance will further enhance the sensitivity of $^{136}$Xe $0\nu\beta\beta$ search.
Furthermore, the PandaX-xT experiment has been proposed to utilize 43-tonne natural xenon in the active volume.
The sensitivity of the effective Majorana neutrino mass is projected to be (10-41)~meV/c$^{2}$, mostly covering the allowed parameter space for the inverted ordering of neutrino mass~\cite{abdukerim2024pandax}. 

\section{Acknowledgment}
This project is supported in part by grants from National Key R\&D Program of China (Nos. 2023YFA1606200, 2023YFA1606202), National Science Foundation of China (Nos. 12090060, 12090062, U23B2070), and by Office of Science and Technology, Shanghai Municipal Government (grant Nos. 21TQ1400218, 22JC1410100, 23JC1410200, ZJ2023-ZD-003). We thank for the support by the Fundamental Research Funds for the Central Universities. We also thank the sponsorship from the Chinese Academy of Sciences Center for Excellence in Particle Physics (CCEPP), Hongwen Foundation in Hong Kong, New Cornerstone Science Foundation, Tencent Foundation in China, and Yangyang Development Fund. Finally, we thank the CJPL administration and the Yalong River Hydropower Development Company Ltd. for indispensable logistical support and other help. 

 \bibliographystyle{elsarticle-num} 
 \bibliography{PandaX_NLDBD}

\clearpage
 \appendix
 \section{Unblinded fit results in four regions of FV.}
 \label{appen}
 \renewcommand{\thefigure}{A}
 \begin{figure*}[hbp]
    \centering
    \subfigure[Middle region]{
        \begin{minipage}[b]{.45\textwidth}
        \centering
        \includegraphics[scale=0.4]{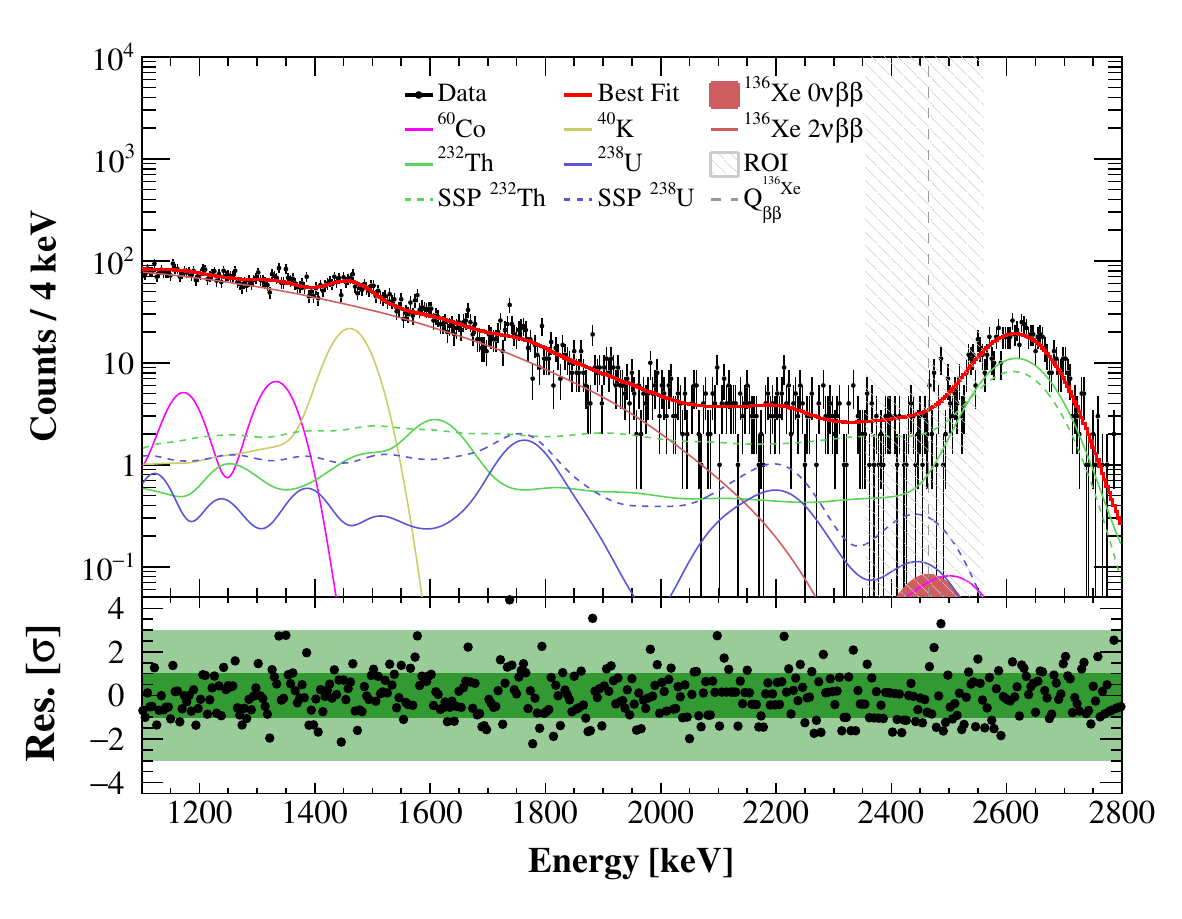}
        \end{minipage}
 }
    \subfigure[Top region]{
        \begin{minipage}[b]{.45\textwidth}
        \centering
        \includegraphics[scale=0.4]{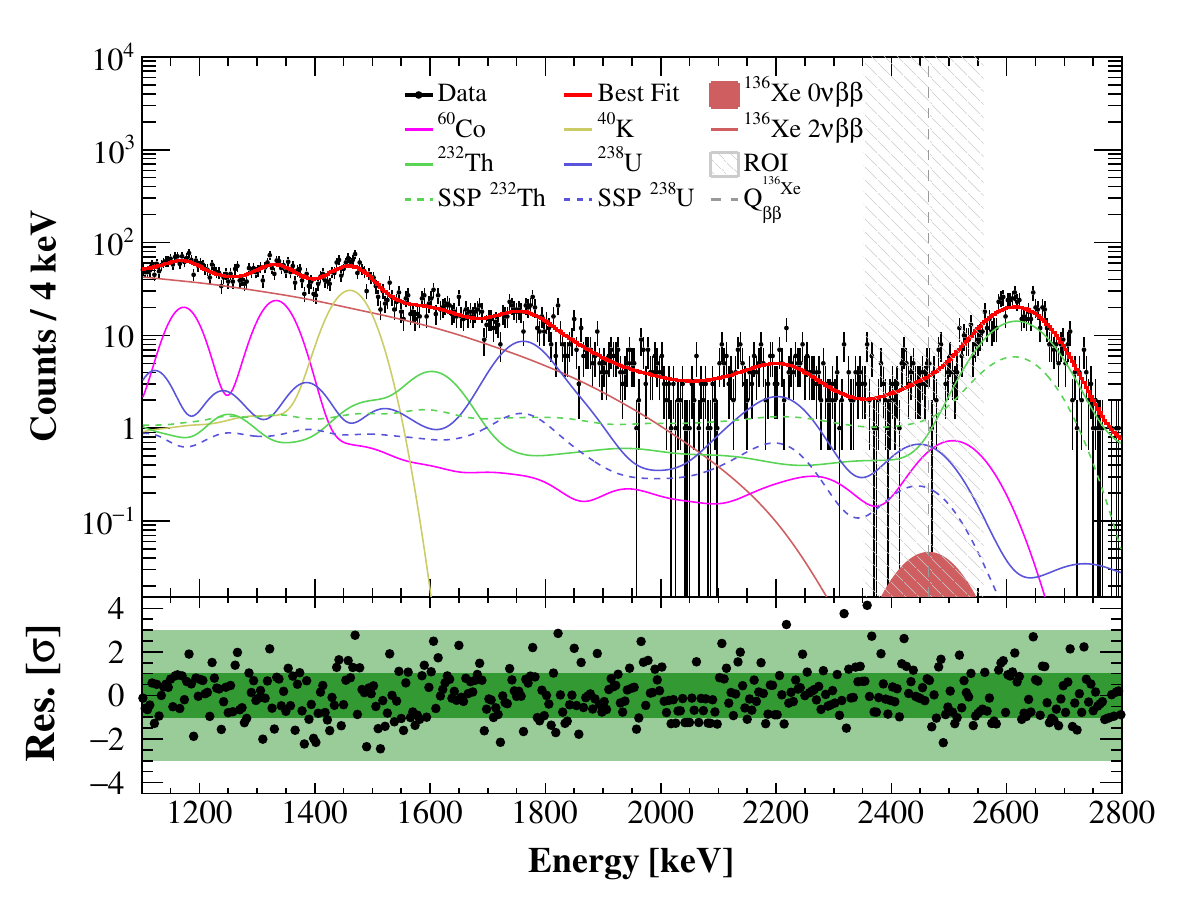}
        \end{minipage}
 }
    \subfigure[Bottom region]{
        \begin{minipage}[b]{.45\textwidth}
        \centering
        \includegraphics[scale=0.4]{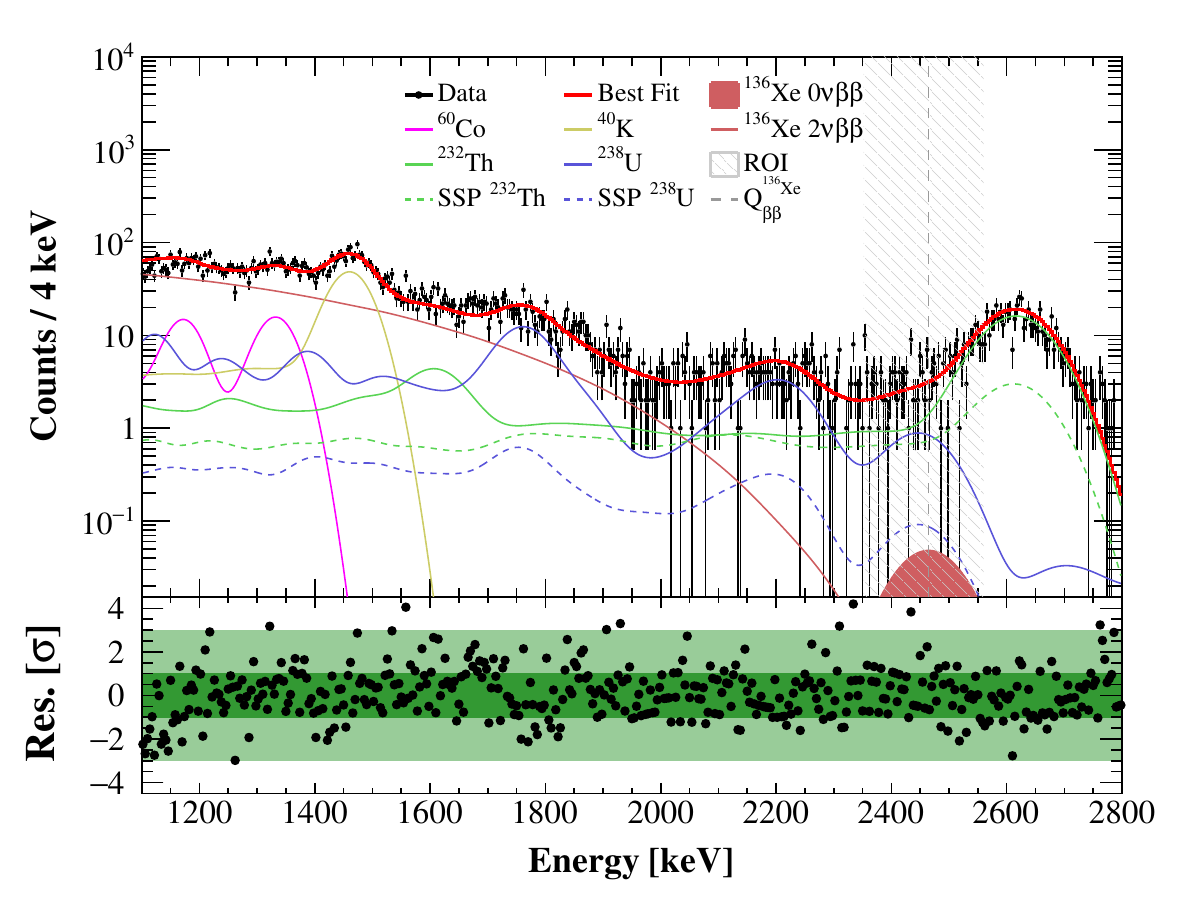}
        \end{minipage}
 }
    \subfigure[Side region]{
        \begin{minipage}[b]{.45\textwidth}
        \centering
        \includegraphics[scale=0.4]{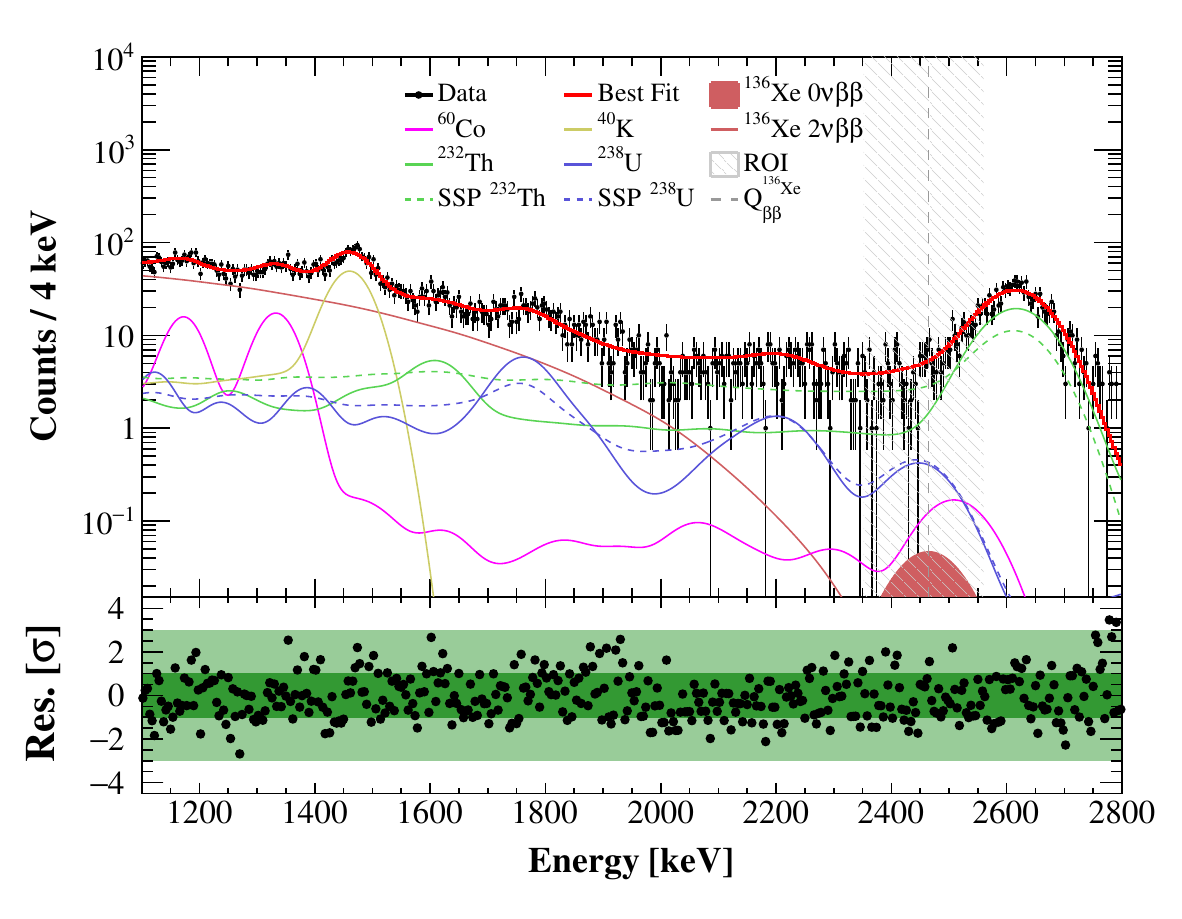}
        \end{minipage}
 }
    \caption{Unblinded fit results in the four regions of FV as indicated in Fig.\ref{fig:FV}. The results of the same region for Run0 and Run1 are combined after fitting for better visualization.}
    \label{fig:4regionFit}
\end{figure*}
\end{document}